\documentstyle[prl,aps]{revtex}
\begin{document}
\draft
\title{\Large\bf  Gauge Invariance,  Finite Temperature
 and Parity Anomaly in D=3 }
\author{S. Deser$^\dagger$, L. Griguolo$^\ddagger$ and 
        D. Seminara$^{\dagger}$}
\address{\it $^\dagger$ Department of  Physics, Brandeis University,
                        Waltham, MA 02254, USA\\
             $^\ddagger$ Center for Theoretical Physics, Laboratory
                         for  Nuclear Science and Department of Physics,\\
                         Massachusetts Institute of Technology, Cambridge,
                         Massachusetts 02139, U.S.A.}
\date{Received \today}
\maketitle
\medskip
\font\ninerm = cmr9
\begin{abstract}
\ninerm\noindent
The effective gauge field actions generated by charged fermions  in $QED_3$ 
and $QCD_3$ can be made invariant  under both small and  large gauge transformations
at any temperature by suitable  regularization of  the  Dirac operator  
determinant, at the price of parity anomalies.
We resolve the paradox that the perturbative expansion is not invariant,
as manifested by the temperature dependence of the induced  
Chern-Simons term, by showing that large ( unlike small) transformations 
and hence their Ward identities, are 
not perturbative order-preserving. Our results are illustrated
through concrete examples of field configurations.
\end{abstract}
Three-dimensional gauge theories are of physical 
interest in the condensed  matter context \cite{CondMatt}, but
display  special features requiring  understanding different from 
their four dimensional counterparts. In particular, we will
be concerned with the complex of problems associated  with the presence of
Chern-Simons  (CS) terms \cite{CS}, the necessary quantization of their
coefficients \cite{CS,Quant}
in the action stemmming from the possibility of making homotopically
nontrivial
{\it ``large"} gauge transformations, and the
effect
of quantum loop corrections on this sector \cite{loop,Dunne1,Pisarski}.
While large transformations  are
always relevant in the nonabelian case, they also come into play in the
physically most interesting case of $QED_3$ at finite temperatures where
the compactified euclidean time/temperature provides a nontrivial, 
$S^1$, geometry.
These exotic features have been the subject of a large literature
\cite{paradox}, as they seemingly lead to a paradox: 
on the one hand, large gauge invariance appears to require
quantization of the CS term's coefficient; on the other, matter loop
contributions to the effective gauge field action at finite temperatures
yield a perturbative expansion in which it acquires temperature-dependent, 
hence non-quantized, coefficients that seems to signal a gauge anomaly. This 
is particularly puzzling  since both the matter action and the process of
integrating out its excitations should be intrinsically gauge invariant.
We will establish that the effective action is indeed invariant under both
small and large transformations using the classic results of \cite{Gilkey}
that gave a clear definition of  the Dirac operator's functional determinant 
by means of $\zeta-$function regularization. Instead, we will see that 
it is the perturbative
expansion that is non-invariant  because large transformations
necessarily introduce  non-analytic dependence on the charge so that
expansion in $e^2$ and large gauge invariance are mutually incompatible: the
induced Chern-Simons term's non-invariance is precisely 
compensated by further
non-local contributions in the effective action. We will also note
the necessary clash between gauge invariance and parity conservation,
similar to that in the familiar axial anomaly in even dimensions. 
All these features  are illustrated in detail
by explicit consideration of some non-trivial configurations 
that enables us  to ``parametrize'' the Chern-Simons aspects
in both the abelian and non-abelian context.

Let us begin with the peculiar properties of large   gauge
transformations that invalidate the usual Ward identity 
consistency. For $U(1)$ in particular, and  restoring explicit 
dependence on $e$, we have
$A_\mu\to A_\mu+e^{-1}\partial_\mu f$. Normally, we can merely
redefine $\tilde f=e^{-1} f$. This is also true at
finite temperature for the small gauge transformations since $f$
is only required to be periodic in Euclidean time $\beta=(\kappa T)^{-1}$.
Thus  a perturbative expansion will be small gauge
invariant  order by order. But for large ones, the periodicity condition becomes
$
f(0,{\bf r})=f(\beta,{\bf r})+2\pi i n$, with  $n\in Z\!\!\! Z$,
and a rescaling will merely hide the $e^{-1}$ factor in the boundary
conditions. This intrinsic dependence means  that only the
{\it full} effective action (which we will show to be invariant),
but not its individual expansion terms (including CS parts !) remains
invariant. [Perturbative non-invariance  will also appear for
any  other expansion, that fails to commute  with the above boundary
condition.]
We are therefore driven to a careful treatment of the
induced effective action $\Gamma[A]$ resulting from integrating out
the charged matter, for us massive fermions, according to the usual
relation
$
\exp\left(-\Gamma[A]\right)=\det(i D\!\!\!\!/+i m)
$
where $D_\mu$ is the $U(1)$ covariant derivative.
The extension to $N$ flavors and to the non-abelian case 
will be seen to be straightforward.
Our 3-space has $S^1({\rm time})\times \Sigma$ topology, 
$\Sigma$ being a compact Riemann 2-surface such as a sphere $S^2$ or  
a torus $T^2$,
depending on the desired spatial boundary conditions. We work with a
finite 2-volume in order to avoid infrared divergences  associated with
the continuous spectrum in an open space. Before proceeding, let us see
how assuming gauge invariance   constrains the form of the determinant. 
[To avoid irrelevant spatial homotopies, we shall 
here take $\Sigma$ to be the
sphere.] Because of the existence of the non-trivial $S^1$ cycle we can
construct (besides $F_{\mu\nu}$) the gauge invariant holonomy 
 $\Omega({\bf r})
\equiv\exp\left (i\int^\beta_0 A_0(t^{'},{\bf r})~dt^{'}\right )$. 
$\Omega$ is not a completely independent variable,
as  part of the information carried by it is already present in
$F_{\mu\nu}$: it satisfies the constraint ${\bf \nabla} \Omega=i
\Omega\int^\beta_0~{\bf E}(t^{'},{\bf r})~dt^{'}$,  implying
that $\Omega$ has the form $\Omega=\exp\left(2\pi i a\right)
\Omega_0({\bf E})$, where $\Omega_0({\bf E})$ is a non local functional depending only
on ${\bf E}$ and on the geometry of $S^2$. The new information is encoded entirely
in the constant $a$, the  flat connection. [For  example, the
non trivial behavior of $A_0$ under large gauge transformation is inherited
by $a$: $a\to a+1$.] Therefore the determinant
can be considered as a
function(al) of $F_{\mu\nu}$ and $a$ alone. Large gauge invariance 
implies the separate Ward identity $e^{-\Gamma(a+1, F_{\mu\nu})}=
e^{-\Gamma(a,F_{\mu\nu})}$,
namely periodicity. Then Fourier-expanding and factorizing out the parity 
anomaly contribution, we obtain
\begin{equation}
\label{coppola}
\exp\left(-\Gamma(F_{\mu\nu},a)\right)=
\exp(iS_{CS})\sum_{k=0}^{\infty}\left(\Gamma^{(1)}_k(F_{\mu\nu})
\cos\pi(2 k-\Phi(F)) a+\Gamma^{(2)}_k(F_{\mu\nu})\sin\pi(2 k-\Phi(F)) a
\right ),
\end{equation}
where $\Phi(F)=\displaystyle{\frac{1}{4\pi}\int d^2x \epsilon^{ij} F_{ij}}$ 
is the electromagnetic flux through $S^2$ and $S_{CS}=\displaystyle{
\frac{1}{4\pi}\int (dx) \epsilon^{\mu\nu\rho}A_\mu\partial_\nu A_\rho}$. 
To write this representation of the effective action we have used the 
fact that Chern-Simons action $S_{CS}$ can be rewritten as $\pi a\Phi(F)$ plus a 
functional of $F$ only. [Effectively, we represent the ``large'' aspects
through $S_{CS}$, or $a$,  and the ``small'' ones through $F_{\mu\nu}$.]
As we shall see, the structure exhibited in  
(\ref{coppola}) will be explicitly realized in our examples.

We now return to  the definition of the effective
action. Within our framework, the Dirac operator is a well-defined elliptic
operator \cite{Gilkey} whose determinant can be rigorously 
specified. The 
$\zeta-$function regularization \cite{Hawking}  defines the formal 
product of all the eigenvalues  $\lambda_n$ as
\begin{equation}
\label{3}
\det i\left(D\!\!\!\!/+m\right)=\Pi \lambda_n
\equiv\exp\left(-\zeta^\prime(0)\right),
\ \ \ \ \ \zeta(s)\equiv\sum  (\lambda_n)^{-s}
\end{equation}
with implicit repetition over degenerate eigenvalues. For $s>3$ in $D=3$
\cite{Gilkey}, the above series converges and its analytic extension 
defines a meromorphic function with only simple poles. It is regular
at $s=0$, thereby assuring the meaningfulness of (\ref{3}). A careful
definition of $\lambda_n^{-s}$ is required to avoid ambiguities. We take
it to be $\exp\left (-s \log\lambda_n\right)$ where the cut is chosen to
be over the positive real axis, $0\le\arg\lambda_n<2\pi$, enabling us to
rewrite $\zeta(s)$ in the more convenient form
\begin{equation}
\label{4}
\zeta(s)=\sum_{{\rm Re}~\lambda_n >0}(\lambda_n)^{-s}+ \exp(  -i\pi~s)
\sum_{{\rm Re}~\lambda_n <0}(-\lambda_n)^{-s}.
\end{equation}
Changing the cut only alters the determinant if it intersects the line
${\rm Im} z=m$, in which case the only relevant difference is the 
sign of the exponential in (\ref{4}).   This alternative choice does not affect 
gauge invariance, but
does change the sign of the parity anomaly terms in $\Gamma[A]$ as was noted 
in \cite{Rossini} by more complicated considerations. Once the determinant
of the Dirac operator has been regularized, its full gauge invariance 
reduces to 
that of its eigenvalue spectrum. But small transformations do not affect the
$\lambda_n$ at all, while the large ones merely permute them, as  in 
usual illustrations of index theorems \cite{pop};   every
well-defined symmetric function of the spectrum, such as $\zeta(s)$ and 
hence $\Gamma[A]$, is unchanged.

The price paid for preserving gauge invariance is (as usual !) an intrinsic
parity anomaly, {\it i.e.}, one present even in the limit when the 
explicitly parity violating fermion mass term is absent. [ That the parity can 
be  sacrificed  for gauge  was effectively noted in \cite{Red}.]
Under $P$,
 $\lambda_n\to -\lambda^*_n$  so that $\zeta^P(s)\ne\zeta(s)$. It is easy
to express the parity violating part $\Gamma^{(PV)}[A]=
1/2(\zeta^\prime(0)-\zeta^{\prime P}(0))$ explicitly in terms of the eta
function in this limit ($m=0$).  Here
\begin{equation}
\zeta(s)-\zeta^P(s)=(1-e^{-i\pi s})\left(\sum_{\lambda_n >0}
(\lambda_n)^{-s}-\sum_{\lambda_n <0}(-\lambda_n)^{-s}\right)\equiv
(1-e^{-i\pi s})\eta(s),
\end{equation}
so that  $\Gamma^{(PV)}[A]=i\pi/2 \eta(0)$. At $m=0$, the continuous 
part of $\eta(0)$ is given in closed form by the CS action \cite{pop,pip}; 
being local means it can be removed by a different choice of regularization.
For $m\ne0$ an expansion in powers  of the mass can be presented
\begin{equation}
\label{potta2}
\Gamma^{(PV)}(A)=
\frac{1}{2}\left.\frac{d}{d~s}( \zeta(s)-\zeta^P(s))\right|_{s=0}=
i\frac{\pi}{2} \eta(0)-
i \sum_{k=0}^\infty  (-1)^k\frac{m^{(2 k+1)}}{2k+1}
\eta (2 k+1),
\end{equation}
while the analogous expansion for the parity-conserving part involves
even powers of the mass\footnote{Several remarks about (5) are in order.
${(a)}$ The presence of the  odd 
powers can be understood as a consequence of the  behavior of
the mass term under parity. Instead, the  anomalous contribution
$\eta(0)$ (proportional to  the even, $m^0$, power)  
originates in a compensation between  vanishing   and 
divergent terms. Similarly for the 
parity-preserving part  there are,  besides the even powers,  
two other possible contributions   in $3$ dimensions,
one proportional to $m$ and one to $m^3$, coming from an analogous
compensation. ${(b)}$ In explicit computations, the expansion,
like its analog for the parity preserving part, must be treated 
carefully, because, even though gauge-invariant order by order,
the coefficients of such expansions are not continuous functional
of the gauge field. [Recall, for example, that $\eta(0)$ jumps by
$\pm 2$ when an eigenvalue crosses  zero or see the Im$\Gamma[A]$
form in the example below.] The total effective 
action is, instead, a continuous functional. ${(c)}$ It would be interesting 
to compare our mass expansions with the one  presented in \cite{pip}, 
obtained from low and high temperature limits in four dimensional 
gauge theories.}. 

For  concrete illustrations of  how the perturbative non-invariance 
paradox is circumvented,
let us now consider some explicit examples of actions and large gauge 
transformations both in the abelian and non abelian sectors. The simplest is 
the pure $S^1$ $(0+1)$dimensional toy model of \cite{Dunne}, with
Dirac operator 
$\displaystyle{\left( i\frac{d}{d t}+A(t)+i m\right )}$ and large
transformations  obeying $f(\beta)-f(0)=2\pi n$.
Charge conjugation $A\to-A$ plays the role of parity, which is violated by
$m$, all as in $(2+1)$. Both the eigenvalues and $\zeta(s)$  
can be obtained exactly in terms
of the average $a=\displaystyle{\frac{1}{2\pi}\int^\beta_0 A(t) dt}$.
We give only the final result here, for $N$ charged fermions:
\begin{equation}
\label{6}
\exp\left(-\Gamma(A)\right)=
\left[2\left(\cosh\left(\frac{\beta m}{2}\right)\cos{\pi a}
-i \sinh\left(\frac{\beta m}{2}\right)\sin{\pi a}\right )
\exp\left(i\pi  a-\frac{\beta m}{2}\right)\right ]^N
\equiv\left (\exp(-\beta m+2\pi i a)+1\right)^N.
\end{equation}
Note that with our regularization,  the action depends on $a$ only 
via the $S^1$ holonomy $\exp(2\pi i a)$.
Expanding (\ref{6}) in terms of $\sin k\pi a$ and $\cos k\pi a$ shows the
consistency of this result with the general  expression (\ref{coppola}). 
A large transformation, $a\to a+1$, leaves (\ref{6}) 
invariant for any $N$, even
or odd, through a sign cancellation between the separate factors 
in the middle term. Note
the necessary presence of an ``intrinsic" charge conjugation anomaly even
at $m=0$: ${\rm Im}\Gamma[A]=iN(a-[a]) $. This is what allows us to preserve
large gauge invariance independently of $N$. Had we opted instead 
(as in \cite{Dunne}) for the
$(0+1)$ equivalent of the more usual, parity-preserving, 
(here $C-$preserving) regularization
the $\exp(i N\pi a)$ factor 
would have been missing and only even $N$ would have kept invariance.
The nonabelian $(0+1)$ scheme is not instructive,
essentially because there  is no equivalent of the abelian CS $\int A$.

A more realistic, $(2+1)$, example is the $U(1)$ field   
\begin{equation}
\label{7}
A_\mu (t,{\bf r})\equiv\left(\frac{2\pi}{\beta} a , {\bf A}({\bf r})\right ),
\end{equation}
where $a$ is a flat connection along $S^1$. ${\bf A}$ lives on
$\Sigma$, with non-vanishing, necessarily integer, flux $\Phi(F)=
n.$ We concentrate on large
transformations  $a\to a +1$, although in
higher genus $\Sigma$ one could also have large trasformations affecting
${\bf A}$. Because of the time independence, we have a tractable eigenvalue 
equation for $\lambda_n$. After some work, it follows that the effective 
action factorizes into two  $(0+1)$ dimensional
contributions like (\ref{6})
and a reduced  expression depending on ${\bf A}$, 
$\Sigma$ and the holonomy  $\exp(2\pi ia)$,
\begin{eqnarray}
\label{bilba}
&&\exp(-\Gamma(A))=
\left [\exp(-\beta m+2\pi i a)+1\right]^{\nu_+}
\left [\exp(-\beta m-2\pi i a)+1\right]^{\nu_-}\\
&&\left |\prod_{\mu_k}\left(1+\exp\left(-\beta\sqrt{\mu^2_k+m^2}
+2\pi  i a \right)\right)\right |^2\ 
\exp\left[2\pi ~\zeta_{\frac{\beta^2}{4\pi^2}
({/\!\!\!\!{\hat D}}^2+m^2)}(-1/2) - (\nu_+ +\nu_-) m \beta\right ].
\nonumber
\end{eqnarray}
Here 
${/\!\!\!\!{\hat D}}$ is the reduced Dirac operator on $\Sigma$, 
$\mu_k$ its nonvanishing eigenvalues\footnote{A simple field configuration 
for which even  the  $\mu_k$ can be 
computed  explicitly is the instanton on the flat unit torus: $A_i=-{\pi n} 
\epsilon_{ij}x^j$. Here $\mu_k^2={4\pi}|n~ k|$ with degeneracy
$2 n$, while $2\pi\zeta_{\frac{\beta^2}{4\pi^2}
({/\!\!\!\!{\hat D}}^2+m^2)}(-1/2)=n\left({4\pi n}
\right)^{1/2}\beta~\zeta_H\left (-1/2,\frac{m^2}{2\pi n}\right)
-(\nu_+ +\nu_-) m \beta$;
$\zeta_H$  is the Hurwitz function.}. The number  of 
positive/negative  chiral zero-modes $v_{\pm}$ of ${/\!\!\!\!{\hat D}}$ 
is represented by $\nu_{\pm}$, with the conventions 
$(\gamma_5\mp 1)v_{\pm}=0$, and the (parity odd) flux is just
$\nu_- -\nu_+$. [ In $(0+1)$ dimensions, there is no chirality,
but an ``opposite  sign'' holonomy can be artificially
introduced by considering also fermions subject to a ``conjugate''
Dirac operator $(-i d/dt-A(t) +im)$ which would change the sign 
of $2\pi i a$ in the last equality of (6).]
That the infinite product in (8)
is convergent  follows from the fact that $\mu_k\simeq c\sqrt{|k|}$ 
\cite{Gilkey}. The  invariance of (\ref{bilba}) under $a\to a+1$ is 
manifest and its structure is consistent with (\ref{coppola}). 
It is clear that a perturbative (i.e., in power of $a$)
expansion of (\ref{bilba}) loses
periodicity in $a$ and hence does not see large invariance order by 
order. For example the Chern-Simons term  ( $S_{CS}=\pi a n$) has a 
coefficient $1-\tanh\left(\frac{\beta m}{2}\right )$. 
The usually quoted coefficient omits the $1$ that represents the 
intrinsic parity-anomaly price of our gauge-invariant regularization and hence
persists at $m=0$. There is actually an ambiguity in its sign (reflecting
the choice of cut in (3)), also present in other regularizations,
for example through the factor $\lim_{M\to\pm\infty} {\rm sign}(M)$ in Pauli-Villars.
 
The analogous finite temperature  ``problem'' 
arises  in the context of the non-abelian 
theory  \cite{Pisarski} as well.  At zero temperature  the loop
correction preserves the integer nature of the Chern-Simons 
coeffficient \cite{Dunne1}, but at finite temperature a puzzling 
temperature dependence  appears \cite{Pisarski}. However
the general discussion presented above can be shown to extend 
naturally to the 
non-abelian case, assuring the gauge invariance of the action.  
To illustrate this, consider the simplest non-abelian generalization 
of the $U(1)-$instanton field considered above: a covariantly 
constant magnetic $SU(2)$ field $F^b_{ij}=\epsilon_{i j}f^b$
on $S^1\times T^2$, whose gauge potential is
$
A^b_\mu\equiv\left(\frac{2\pi}{\beta} a,
-{\pi n}\epsilon_{ij}x^j \right)f^b, 
$
where $f^b$ is a unit color vector  and $n$ an integer. 
The relevant mechanism here is actually quite different from the abelian
case. There the spectral asymmetry entailing the parity anomaly was 
governed by the flux $\Phi(F)$ on $\Sigma$:  geometrically $\Phi(F)$ 
represents  a nonvanishing Chern class for the reduced 
2-dimensional field. But  the Chern class of a $D=2$ 
non-abelian gauge field vanishes:  the 
asymmetry of the spectrum is not due to the difference in chirality of 
the zero-modes of the reduced Dirac operator on  $T^2$ (the kernel 
being chirally symmetric) but rather to their  different structure as 
multiplets of $SU(2)$.  Consequently the determinant  yields the 
abelian result, with $\nu_\pm$ replaced by $2 \nu_\pm$. 
To see this, imagine  aligning   $f^b$ along  say the $3-$direction.
Then the eigenvalue problem splits into two $U(1)$'s  coupled respectively
to $\pm A$, so that we just get a doubling of the one-component 
abelian result. [For $SU(N)$, one would align  $f^b$ along the 
Cartan sub-algebra, thereby again splitting into various abelian sectors,
with different charges, in a well-defined way.]
In this non abelian context,   the general characteristics we have 
considered here such as  parity anomalies and large gauge-invariance 
persist at zero temperature and have been discussed, 
with explicit examples in \cite{forte}

In conclusion, we have shown that the apparent large gauge anomalies resulting 
from a perturbative expansion of the full effective action are due to  the more 
complicated (order-violating) nature of the Ward identities when a non-trivial
homotopy is  present, the action itself being fully gauge invariant with 
suitable regularization, one that necessarily entails parity anomalies. 
This has been  illustrated by explicit abelian
and non abelian field configurations. Details will be given elsewhere.

\medskip
This work is supported by NSF grant PHY-9315811, 
in part by funds provided by the U.S. D.O.E.
under cooperative agreement \#DE-FC02-94ER40818
and by INFN, Frascati, Italy.

\end{document}